\documentclass[12pt]{iopart}

\usepackage{graphicx}
\usepackage{bm}

\begin{document}

\title{Activity ageing in growing networks}

\author{ R. Lambiotte$^1$ }

\address{
$^1$ Universit\'e de Li\`ege, Sart-Tilman, B-4000 Li\`ege, Belgium
}
\ead{renaud.lambiotte@ulg.ac.be}

\begin{abstract}
We present a model for growing information networks  where the ageing of a node depends on the time at which it entered the network and on the last time it was cited. The model is shown to undergo a transition from a small-world to large-world network. The degree distribution may exhibit very different shapes depending on the model parameters, e.g. delta-peaked, exponential or power-law tailed distributions.
\end{abstract}



\maketitle

\section{Introduction}

The ageing of nodes 
is an important process in order to understand the way information or social networks grow \cite{ageing0,ageing,hajra,hajra2,lehm,yule}. For instance, this process may be responsible for deviations to scale-free degree distributions \cite{ageing0} or for the non-vanishing values of the clustering coefficient observed in many networks \cite{klemm,klemm2,tian}.
Ageing accounts for the fact that {\em old} nodes lose their ability
to acquire new links as time goes on, thereby limiting the number of active nodes to a small fraction of the whole network. In general, this effect embodies the notion of generation  for social agents, the lifetime of an information or of an article, etc... Such effects may be taken into account by attributing  an age $\tau$ to nodes \cite{ageing} and by assuming that their probability to receive a link from a newly entering node depends on their age (through some decreasing function of $\tau$) and, possibly, on other parameters such as their degree $k$ (preferential attachment \cite{bara1,kra1}). An alternative model \cite{klemm,klemm2,tian} assumes that nodes can be deactivated with a probability proportional to $k^{-1}$. In this deactivation model (DM), once a node is deactivated, it is excluded from the network dynamics. DM is appealing because it mimics the fact that less popular nodes are more easily {\em forgotten} than the popular ones. This is  the case for citation networks \cite{redner} (e.g. nodes are the articles and directed links are the citations of one article by another one), for instance, where 
 highly cited papers usually continue to be cited for a long time, and vice 
versa. E.g.   
papers with more than 100 citations have an average citation age of
$11.7$ years while  the  publications with more than 
1000 citations have average citation age of $18.9$ years \cite{redner}.
Unfortunately, DM is unsatisfactory because the underlying mechanism for this deactivation probability $\sim k^{-1}$ is not identified and a more fundamental model is therefore of interest.

A similar lack of clarity also occurs when one tries to justify linear preferential attachment models \cite{bara1,price}. Indeed, the latter imply that entering nodes have a global knowledge of the network, i.e. they must be aware of the degrees of every 
previously existing nodes before connecting to one of them. This unrealistic approach can be elegantly circumvented by introducing redirection \cite{krapi,dir} or copying \cite{GNC,valverde,japan,protein} mechanisms. In the simplest version, that one explains in terms of citation networks for the sake of clarity, 
 an author who is writing the reference list for a new paper 
picks a random pre-existing paper. Then the author cites either the randomly selected paper (with probability $1-r$) or one of the 
references within that paper (with probability $r$). It is straightforward to show that this 
purely local process generates linear preferential attachment \cite{krapi}. 
In this Article,  we proceed along the same line of thought and introduce a model, called Link Activation Model (LAM), that includes ageing effects. Its interpretation is quite natural for information networks, such as citation networks. The system is a growing network where, for the sake of simplicity, entering nodes have only one outgoing link (each paper cites one other paper). One assumes that only recent nodes are active but, contrary to previous models, a node is active if it has been introduced recently {\bf or} if its has been cited recently. In detail, when an author cites a paper, it either selects the latest paper (the paper entered at the previous time step) with probability $p$ or a random paper with probability $1-p$. Then, with probability $r$, the author cites the paper cited by the selected paper. With probability $1-r$, he cites the selected paper. The model therefore depends on two parameters $p$ and $r$ that measure the importance of ageing and redirection processes as compared to random effects. There are four different possibilities that can take place at each time step, as summarized in Fig.1. An applet allowing the dynamical visualisation of the model should also be available online \cite{online}. Let us stress that the ingredients of the model are very general and that LAM is not limited to citation networks, but should also apply to other information networks, e.g. the Web.

\begin{figure}
\includegraphics[width=6.3in]{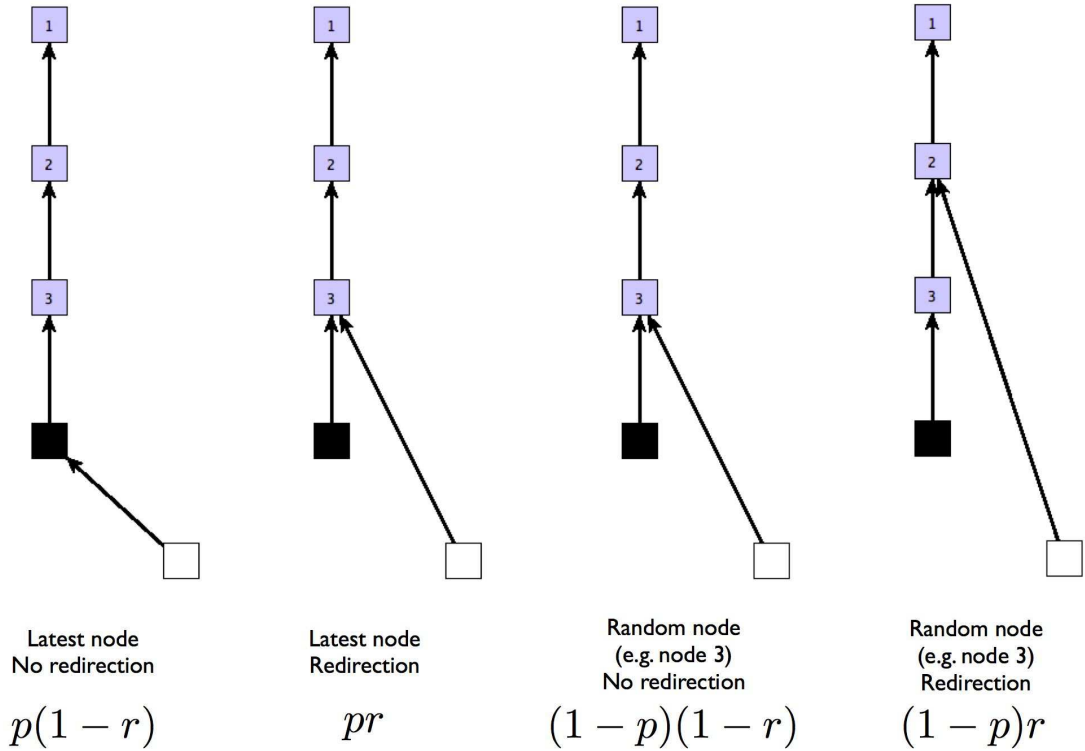}

\caption{Four possible configurations when a new node enters the network. The latest node is darkened and the entering node is in white.  With probability $p (1-r)$, the latest node receives the link from the entering node. With probability $p r$, the latest node is selected, but redirection takes place, so that the {\em father} of the latest node receives the link from the entering node. The two other possible configurations, associated to the random selection of a node (in this example, node 3), occur with probabilities  $(1-p)(1-r)$ and $(1-p) r$.}
\label{fig1}      
\end{figure}

Before going further, let us precise notations. Initially ($t=0$), the network is composed of one node, the seed. For the sake of coherence, the seed has an outgoing link connected to itself. At each time step $t$, a new node enters the network.  Consequently, the total number of nodes is equal to $N_t=1+t$, and the number of links is also $L_t=1+t$.

\section{Height distribution}

In this section, we focus on the height distribution,    
 the height of  a node \cite{bennaim} being defined to be the length of the shortest path between this node and the seed. Let us note $H_{g;t}$ the average number of nodes at the height $g$. By construction, $H_{0;t}=1$ for all times. We also define $l_{g;t}$ to be the probability that the latest node is at height $g$. It is straightforward to show that these quantities satisfy the coupled rate equations

\begin{eqnarray}
H_{g;t+1} &=& H_{g;t} + (1-p) \frac{(1-r) H_{g-1;t} + r H_{g;t}}{t+1} + p [(1-r)l_{g-1} + r l_{g}]\cr
l_{g;t+1} &=& (1-p) \frac{(1-r) H_{g-1;t} + r H_{g;t}}{t+1} + p [(1-r)l_{g-1} + r l_{g}],
\end{eqnarray}
except for $g=1$:
\begin{eqnarray}
H_{1;t+1} &=& H_{1;t} + (1-p) \frac{H_{0;t} + r H_{1;t}}{t+1} + p (l_{0;t} + r l_{1;t})\cr
l_{1;t+1} &=& (1-p) \frac{ H_{0} + r H_{1;t}}{t+1} + p (l_{0;t} + r l_{1;t})
\end{eqnarray}
and for $g=0$ where one has the trivial solutions $H_0=1$ and
$l_0=0$ (this is due to the fact that an entering node can only arrive at height $1$ or higher). The above rate equations are derived in the usual way and generalise the equation with $p=0$ found in \cite{lambi} for instance. It is straightforward to verify that $N_t=\sum_g H_{g;t}=t+1$ and $l_t=\sum_g l_{g;t}=1$.

Let us first focus on the case $p=1$, where only latest nodes are selected, and take a continuous time limit (this is justified a posteriori as we are interested in the long time behaviour of the model). In that case, one has to solve 
\begin{eqnarray}
\partial_t H_{g;t} &=& (1-r)l_{g-1} + r l_{g}\cr
\partial_t l_{g;t} &=& (1-r) l_{g-1} + (r-1) l_{g}.
\end{eqnarray}
 In the following, we are interested in the behaviour of the average total height $G_t=\sum_{g=0}^{\infty} g H_{g;t}$. To do so, one also needs to evaluate the behaviour of $z_t=\sum_{g=0}^{\infty} g l_{g;t}$ which is the average height of the latest node and is easily found to satisfy
\begin{eqnarray}
\partial_t z_{t}= (1-r). 
\end{eqnarray}
Consequently, $z_t$ asymptotically behaves like $(1-r) t$ and the equation for the total height $G_t$ reads
\begin{eqnarray}
\partial_t G_{t} =  (1-r) + (1-r) t.
\end{eqnarray}
This equation  leads to the asymptotic behaviour $G_t = \frac{(1-r)}{2} t^2$. This implies that the average height $g_t\equiv G_t/(N+1) \simeq G_t/t$ asymptotically increases linearly with time. Moreover,  the redirecting process slows down the growth of the network (see Fig.2).  This is expected as redirection favours the connection to nodes closer to the seed. In the limiting case $p=1$, where the process is easily shown to lead to a star network (i.e. all the nodes are connected to the seed), one finds $G_t = t \Leftrightarrow g=1$.

\begin{figure}

\includegraphics[width=6.3in]{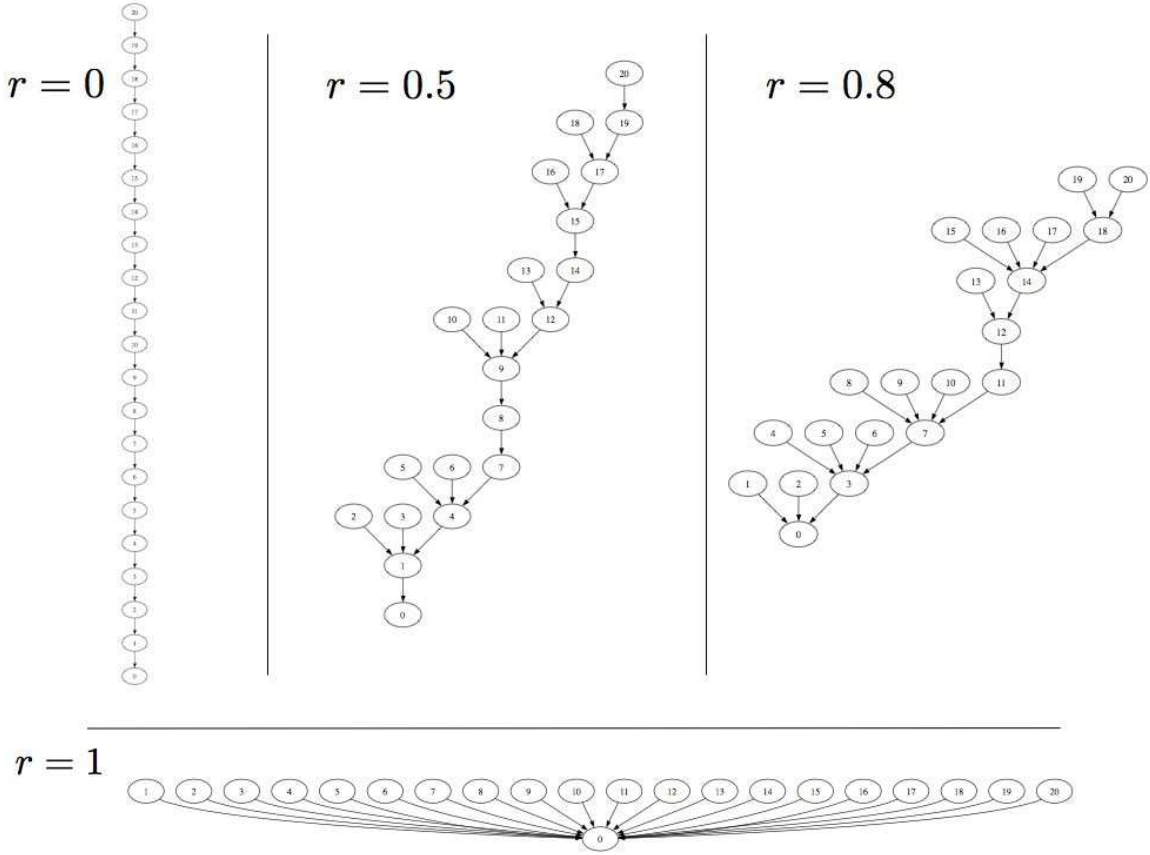}

\caption{Typical realisations of the model when $p=1$. In that case, the average height evolves linearly with time and one observes a large range of behaviours, from a aligned network ($r=0$) to a star network ($r=1$). The average height $g$ increases in a large-world way, i.e. linearly with time $g_t = \frac{(1-r)}{2} t$.}
\label{fig2}      
\end{figure}

Let us now focus on the more general case $p<1$ which reads in the continuum time limit 
\begin{eqnarray}
\partial_t H_{g;t} &=&  (1-p) \frac{(1-r) H_{g-1} + r H_{g;t}}{t+1} + p [(1-r)l_{g-1} + r l_{g}]\cr
\partial_t l_{g;t} &=& (1-p) \frac{(1-r) H_{g-1} + r H_{g;t}}{t+1} + p [(1-r)l_{g-1} + r l_{g}] - l_g.
\end{eqnarray}
By using
\begin{eqnarray}
\frac{H_{0} + r H_{1;t}}{t+1} + \sum_{g>1} g \frac{(1-r) H_{g-1} + r H_{g;t}}{t+1} = r \frac{H_{0}}{t+1}+(1-r) + \frac{G_t}{t+1} ,
\end{eqnarray}
and neglecting terms  $\sim t^{-1}$, one obtains the following set of equations for the above defined average quantities
\begin{eqnarray}
\label{compl}
\partial_t G_t &=&  (1-p)    (1- r  +   \frac{G_t}{t} ) + p (1-r + z_t)\cr
\partial_t z_t &=& (1-p)    (1- r +   \frac{G_t}{t} )+ p (1-r + z_t) - z_t.
\end{eqnarray}
It is easy to simplify Eq.\ref{compl} into:
\begin{eqnarray}
\label{simpl}
t \partial_t G_t &=&    (1- r) t + (1-p)  G_t  + p ~t ~z_t \cr
t \partial_t z_t &=& (1- r) t + (1-p)  G_t  + (p-1) t ~z_t.
\end{eqnarray}
Numerical integration of the above set of equations and our knowledge of the previous simplified cases (e.g. Eq.5) suggest to look for solutions of the
form   $G_t= C t \log(t)$, $z_t=C \log(t)+ K$. By inserting these expressions into Eqs.\ref{simpl} and
keeping leading terms in the long time limit  $t>>1$, one finds the conditions  
\begin{eqnarray}
 C = K = \frac{1-r}{1-p},
 \end{eqnarray}
which cease to be valid when $p=1$, in agreement with the solution of Eq.5. Consequently, the average height $g_t$ asymptotically grows logarithmically with time $g_t = \frac{1-r}{1-p} \log(t)$. This result should be compared with the linear regime $g_t = \frac{(1-r)}{2} t$ taking place when $p=1$.
Let us stress that such a transition from a large-world ($g_t \sim t$) to a small-world \cite{watts,havlin} ($g_t \sim \log(t)$) network has already been observed in another model with ageing \cite{tian} and is associated with the cross-over 
from a structured network, reminiscent of a one-dimensional line, to an unstructured network. The above solution is in agreement with the prediction $g_t = (1-r) \log(t)$ taking place in a model without ageing \cite{lambi}.

\section{Degree distribution}

Let us note by $N_{k;t}$ the average number of nodes with $k$ incoming links. For the sake of clarity, we first focus on three simplified cases, $r=0$, $p=0$ and $p=1$ before deriving results for general values of the parameters.

When $r=0$, there is no possible redirection and the stochastic mechanism takes place during the selection of a node. With probability $p$, the latest node, which has by definition zero incoming links, receives the link of the entering node, while with probability $1-p$, a random node receives this link. Consequently, the rate equation for $N_{k;t}$  \cite{sitges} reads
\begin{equation}
\label{NkN}
\partial_t N_k = (1-p) \frac{N_{k-1}-N_k}{N} + p (\delta_{k,1}-\delta_{k,0})  +\delta_{k,0},
\end{equation}
where the last delta term accounts for the degree distribution of the newly entering node.
We look for a stationary solution of the distribution $n_k=N_k/N$ which is determined by the recurrence relations
\begin{eqnarray}
  (1-p) (n_{k-1}-n_k) + p (\delta_{k,1}-\delta_{k,0})  +\delta_{k,0} - n_k=0.
\end{eqnarray}
Its solution is easily found to be
\begin{eqnarray}
n_0 &=& \frac{1-p}{2-p}\cr
n_1 &=& \frac{1-p}{2-p} n_0 + \frac{p}{2-p} \cr
 n_k &=&  \left(\frac{1-p}{2-p}\right)^{k-1} n_1, ~~~{\rm for}~ k>1.
\end{eqnarray}
When $p=0$, one recovers the exponential solution $n_k=(1/2)^{k+1}$. For increasing values of $p$, the tail of the distribution remains exponential, but its core is more and more peaked around $k=1$. In the limiting case $p=1$, the solution goes to a peaked distribution $n_k=\delta_{k1}$ that  corresponds to an aligned network (see Fig.2).

In the case $p=0$, LAM reduces to the usual model with redirection for which it is well-known \cite{krapi} that the degree distribution evolves as
\begin{eqnarray}
\partial_t n_k = r  [(k-1) n_{k-1} - k n_k] + (1-r) (n_{k-1}-n_k)  +\delta_{k,0}  - n_k .
\end{eqnarray}
The stationary solution is therefore found by recurrence
\begin{eqnarray}
 (  r k +  2 -r) n_k =   (  r k  + 1-2 r) n_{k-1} .
\end{eqnarray}
This stationary solution has a power-law tail $k^{-\nu}$ whose exponent $\nu$ is obtained by inserting the form $n_k \sim k^{-\nu}$ into the above equation. By keeping the leading terms in $k^{-1}$, i.e. $(k-1)^{-\nu} = k^{-\nu} (1-1/k)^{-\nu} \simeq k^{-\nu} (1+\nu /k) $, one has to solve
\begin{eqnarray}
 (  r k +  2 -r) k^{-\nu} &=&   (  r k  + 1-2 r) k^{-\nu} (1+\nu /k),
 \end{eqnarray}
so that one recovers the value $ \nu = \frac{1+r}{r} $ derived in \cite{krapi}.

The case $p=1$ is slightly more complicated, due to the fact that the selected node is always the latest node. Consequently, one also has to focus on the quantity $A_k$ that is the average number of nodes with degree $k$ that are cited by the latest node. By construction, this quantity satisfies $\sum_k A_k=1$ (because there is only one latest node by construction and this latest node has only one outgoing link) and the system is described by the coupled set of equations
\begin{eqnarray}
\label{NkN2}
\partial_t N_k &=& r (A_{k-1}-A_k) + (1-r) (\delta_{k,1}-\delta_{k,0})  +\delta_{k,0} \cr
\partial_t A_k &=& r A_{k-1} + (1-r) \delta_{k,1} - A_k .
\end{eqnarray}
Let u note that the equations for $N_k$ and $A_k$ are quite similar, except for their loss term. This is due to the fact that all nodes that do not receive a link at a time step are cited by nodes that cease to be  the latest node by construction.
The stationary values of $A_k$
\begin{eqnarray}
A_0 &=& 0 \cr
 A_k &=& r^{k-1} (1-r), ~~~ {\rm for}~ k>0
\end{eqnarray}
and of  the distribution $n_k$
\begin{eqnarray}
n_0&=&r \cr
n_k&=& r^{k-1} (1-r)^2, ~~~ {\rm for}~ k>0
\end{eqnarray}
are found by recurrence. In the case $r=0$, one recovers the distribution $n_k=\delta_{k1}$ of the aligned network. Before going further, let us stress that LAM exhibits a very rich phenomenology, with a degree distribution that can behave like a delta peak, an exponential or a power-law depending on the parameters. 

By putting together the contributions of  the above limiting case, it is straightforward to write a set of equations for general values of $p$ and $r$:
\begin{eqnarray}
\label{NkN3}
\partial_t  N_k  &=&  (1-p) [r  \frac{(k-1) N_{k-1} - k N_k}{N} + (1-r) \frac{N_{k-1}-N_k}{N}] \cr
&+&  p 
[ r (A_{k-1}-A_k) + (1-r) (\delta_{k,1}-\delta_{k,0})]  +\delta_{k,0} \cr
\partial_t A_k  &=&  (1-p) [r  \frac{(k-1) N_{k-1} }{N} + (1-r) \frac{N_{k-1}}{N}] \cr
&+& p [r A_{k-1} + (1-r) \delta_{k,1}]  - A_k ,
\end{eqnarray}
whose stationary solutions are found by resolving the recurrence relations
\begin{eqnarray}
\label{rec}
0&=& (1-p) [r  ((k-1) n_{k-1} - k n_k) + (1-r) (n_{k-1}-n_k)]\cr  &+& p 
[ r (A_{k-1}-A_k) + (1-r) (\delta_{k,1}-\delta_{k,0})]  +\delta_{k,0}  - n_k \cr
0 &=& (1-p) [r  (k-1) n_{k-1}  + (1-r) n_{k-1}] \cr
&+&  p [r A_{k-1} + (1-r) \delta_{k,1}] - A_k.
\end{eqnarray}
It is possible to write the formal solution of the second relation:
\begin{eqnarray}
A_0&=& 0 \cr
 A_k &=& (p r)^{k-1}  p (1-r)  + \sum_{i=1}^{k} (p r)^{i-1}  (1-p) [r  (k-i) + (1-r)] n_{k-i}.
\end{eqnarray}
After inserting this solution into the first equation of Eqs.\ref{rec}, looking for a solution of the form $n_k \sim k^{-\nu}$ and keeping the leading terms in $k^{-1}$, it is straightforward but lengthy to get the expression
 \begin{eqnarray}
 \label{solu}
 \nu=\frac{1+r-2 p r }{r-p r} .
\end{eqnarray}
This solution is well defined when $p \neq 1$ and $r \neq 0$ and recovers the result derived above when $p=0$. It is  important to note that the tail of the distribution behaves like a power-law for any other value of the parameters. Let us also note that Eq.\ref{solu} is a monotonically increasing function of $p$, for fixed values of $r$, so that ageing mechanisms have a tendency to diminish the number of nodes with very high degrees. This can be understood by noting that ageing diminishes the probability for old nodes to be cited, while these old nodes are typically those with the highest degree.

\section{Discussion}
In this Article, we have presented a simple model for growing networks with ageing. This Link Activated Model incorporates the fact that articles remain present in the {\em collective memory} as long as they are cited or read. Namely, articles that are the most likely to be cited are those that have been published recently or those that have been cited recently. In other words, all sorts of articles that are  {\em present} and may have punctually triggered the  reader's curiosity.
This natural process is shown to lead to a rich behaviour for the network structure, that leads to  a transition from a large-world to a small-world network. Moreover,  various kinds of asymptotic stationary degree distributions may be reached depending on the model parameters: a delta peak that corresponds to a one-dimensional lattice, exponential-like distributions or power-law tailed distributions. Let us insist on the fact that LAM is quite general and should apply to many situations involving a competition between multiplicative effects ({\em rich gets richer}) and ageing. Apart from citation networks that have been discussed above, one may think of {\em short-lived} information web-pages. A typical example is {\em digg.com}  where users propose a new information/article and are subject to the votes of the whole community of users. Usually, informations lose their appeal within a few hours or days. 

 {\bf Acknowledgements}
This work has been supported by European Commission Project 
CREEN FP6-2003-NEST-Path-012864. I would like to thank J.-P. Boon for fruitful comments.

\end{document}